\documentclass[apj]{emulateapj}
\usepackage{graphicx}
\shortauthors{Ro and Matzner}
\begin{document}
\title{Shock Emergence in Supernovae: Limiting Cases and Accurate Approximations}
\author{Stephen Ro \& Christopher D. Matzner}
\affil{Department of Astronomy \& Astrophysics, University of Toronto, 50 St. George St., Toronto, ON M5S 3H4, Canada}
\date{\today}
\begin{abstract}
We examine the dynamics of accelerating normal shocks in stratified planar atmospheres, providing accurate fitting formulae for the scaling index relating shock velocity to the initial density and for the post-shock acceleration factor as functions of the polytropic and adiabatic indices which parameterize the problem.   In the limit of a uniform initial atmosphere there are analytical formulae for these quantities.  In the opposite limit of a very steep density gradient the solutions match the outcome of shock acceleration in exponential atmospheres.   
\end{abstract}
\keywords{hydrodynamics -- shock waves --- supernovae: general} 

\section{Introduction}


Shock emergence at the surface of an exploding star is an important moment in the life of a supernova.  Shock and post-shock acceleration in the outer stellar envelope, and the breakout of post-shock radiation from a thin layer beneath the photosphere, can have a number of significant consequences.   The escaping flash of radiation gives an energetic precursor which can signal the supernova's existence \citep{1978ApJ...223L.109K} and carries physical information about the explosion \citep{1999ApJ...510..379M, 2004MNRAS.351..694C,2012ApJ...747...88N,2010ApJ...725..904N,2011ApJ...742...36S,2010ApJ...716..781K,2010ApJ...717L.154S,2010ApJ...708..598P}; traveling outward, it can ionize a circumstellar nebula like the one surrounding SN 1987A \citep{1996ApJ...464..924L} and produce an infrared light echo as it encounters dust  \citep{2008ApJ...685..976D}.    Shock emergence launches the fastest ejecta, the first to host the supernova photosphere \citep{1992ApJ...394..599C} and the first to interact with circumstellar and interstellar matter, producing a synchrotron-emitting shell \citep{1996ApJ...461..993F}.   If they meet a companion star or dense circumstellar disk, an additional x-ray signal can be produced \citep{2010MNRAS.409..284M,2010ApJ...708.1025K}.

 In particularly compact and energetic explosions the shock can become relativistic before emerging, and relativistic ejecta can create x-ray and $\gamma$-ray transients in their circumstellar collisions  \citep{1999ApJ...510..379M,2001ApJ...551..946T}, and may produce light elements through spallation 
\citep{2002ApJ...581..389F,2004ApJ...610..888N} and, potentially, ultra-high-energy cosmic rays \citep{2007PhRvD..76h3009W,2008ApJ...673..928B}.   

Potentially observable shock breakouts accompany several other types of astrophysical events, including the type Ia explosions \citep{2010ApJ...708..598P} and accretion-induced collapses  \citep{1999ApJ...516..892F,2001ApJ...551..946T} of white dwarfs, tidal disruptions of stars, jet and cocoon emergence in long-duration gamma-ray bursts, and (albeit in a less energy-conserving manner) superbubbles in galactic disks. 

Underlying all these phenomena are the hydrodynamics of shock acceleration in the outer layers of a star, and anchoring these dynamics is the asymptotic problem of flow behind a normal, adiabatic shock accelerating through a planar medium which varies as a power law with depth.   As \citet{1999ApJ...510..379M} first demonstrated,  this asymptotic planar solution can be combined with the dynamics of a spherical, self-similar blastwave into an accurate approximate model for shock propagation and post-shock flow in a spherical explosion.  This, in turn, can be used to predict the amount and upper speed limit of the fastest ejecta and properties of the breakout flash \citep{1999ApJ...510..379M,2004MNRAS.351..694C}, transition to relativistic flow and aspects of the circumstellar interaction \citep{2001ApJ...551..946T}, and many other breakout-related phenomena.   With advances in the theory of photon-mediated shocks and emission around the time of breakout \citep[e.g.,][]{2010ApJ...716..781K,2010ApJ...725..904N,2011ApJ...742...36S,2012ApJ...747..147K},  of the ultra-relativistic self-similar problem \citep{2002ApJ...569L..47P,2005ApJ...627..310N,2006ApJ...643..416P,2007ApJ...657..860K}, and of the interaction of relativistic ejecta with a stellar wind \citep{2006ApJ...645..431N}, among others, there are ample opportunities for these approximate global models to be improved and extended.   

To advance this larger project we focus here on the planar, adiabatic, non-relativistic problem of an accelerating normal shock.   Our goal is to provide flexible yet highly accurate approximations for the most important flow quantities, the shock acceleration index and the post-shock acceleration factor, as functions of the adiabatic and polytropic indices ($\gamma$ and $\gamma_p$, respectively) which parameterize the problem.  A secondary goal is to demonstrate that although the flow quantities must typically be found as eigenvalues of the dynamical problem, they adhere to well-understood limiting forms in several asymptotic cases. 

The self-similar problem with a power-law atmosphere below vacuum was posed by \citet{1956SPhD....1..223G} and solved in its Eulerian form by  \citet{Sakurai60}.   We shall use \citeauthor{Sakurai60}'s eigenvalue method to identify the shock acceleration index, but for the post-shock flow we employ the Lagrangian approach by \citet{1999ApJ...510..379M}.  This has the dual advantage that it continuously describes both the pre-breakout and post-breakout flow in a single function, and that it naturally connects each fluid element's state at the shock front with those in the final state.

\section{Problem, method, and solutions}\label{S:Method}
Our problem involves one dimensional flow with an altitude $x$ relative to the stellar surface.  The initial density distribution of cold matter is $\rho_0(x) \propto (-x)^n$ for $x<0$ and $\rho_0(0)=0 $ for $x>0$.  Here $n$ is the polytrope parameter, which is related to $\gamma_p$ by the hydrostatic relation with constant gravity $g_*$: if $P=0$ at $x=0$, then $P(x) = g_*(-x)\rho(x)/(n+1) \propto \rho(x)^{\gamma_p}$ with $\gamma_p = 1+1/n$.    A strong adiabatic shock wave accelerates down this density gradient, reaching $x=0$ at $t=0$ with infinite velocity;  neglecting radiative effects, this is the point of breakout.  For $t>0$ the shock disappears and matter expands into the region of positive $x$.   In the limit that all additional physical effects -- curvature, gravity, and temperature of the star, finite depth, non-simultaneity of breakout, relativity, shock thickness, etc. -- are negligible, the flow is self-similar.  The shock velocity accelerates according to $v_s = \dot x_s(t)  \propto (-x)^{-\lambda} \propto \rho^{-\beta}$, where $\beta = \lambda/n$.   The fluid motion is a universal function of self-similar variables like $x/x_s(|t|)$ or $x(m,t)/x_0(m)$, in which each fluid element (labeled by its mass coordinate $m$) accelerates from a post-shock velocity toward its terminal velocity, which is a unique multiple $v_f(m)/v_s(m)$ of the shock velocity which crossed that element.  Our task is to find $\beta$ and $v_f/v_s$ as functions of $\gamma$ and $\gamma_p$.

\subsection{Shock acceleration parameter and its limits}\label{SS:lambda_and_beta}
To find the shock acceleration index $\lambda$, or equivalently the velocity-density index $\beta$, we follow \citet{Sakurai60}.   \citeauthor{Sakurai60} writes the conservation equations for mass, energy, and entropy in Eulerian form, introduces the self-similar ansatz, and arrives at a single, first-order differential equation for the spatial structure of the post-shock flow prior to breakout.   This equation must pass smoothly from the conditions immediately behind the shock front, through a critical point at the sonic point of the flow; this is only possible for a unique value of $\lambda$ or $\beta$, which we identify by a shooting method.   We present the solution space $\beta(n,\gamma)$ spanning $\log_{10}n=-6,..., 6$ and $\log_{10}(\gamma-1)=-5,...,6$ in Table \ref{Table:beta}.  For the entire parameter space, the simple functional fit 
\begin{eqnarray} \label{betaFit} 
\beta &=& { \left[ A + \left( B \gamma \over \gamma-1\right)^C\right]^{-1} } ~~\mbox{with}\\ 
A&=& 2-{0.22 \over 0.59 n^{-9/8} + 1}, \nonumber \\ 
B &=& 2 + {2.31 \over 1.72 n^{-1} + 1}, ~~\mbox{and}\nonumber \\ 
C &=& \frac12 - {0.0312 \over 1.1 n^{-1} +1} \nonumber 
\end{eqnarray} 
to be quite accurate: {the root-mean-square error relative to the values in table \ref{Table:beta} is 1.4\%, and the error, which is concentrated at high $n$ and low $\gamma$, is at most 4.0\%.    High accuracy is necessary, because $\beta$ is the exponent of a number which becomes large around breakout.  For instance, the energy in relativistic ejecta scale as $E_{\rm rel}\propto [E_{\rm in}/(M_{\rm ej}c^2)]^{\gamma_p/(2\beta)}$, where $E_{\rm in}$ is the explosion energy and $M_{\rm ej}$ is the total ejected mass; in the model for SN 1998bw discussed by \citet{2001ApJ...551..946T}, an error of $\beta$ leads to an error in $E_{\rm rel}$ which is nine times greater.   Higher accuracy can be obtained by interpolating our table, and the differential equations yield solutions to numerical accuracy. } Several limiting forms of our fit to $\beta(n,\gamma)$ are readily apparent.  

In the limit $n\rightarrow 0$ of an effectively uniform stellar envelope, $\beta$ and its fit in equation (\ref{betaFit}) reproduce the approximate expression derived by \citeauthor{whitham58} (\citeyear{whitham58} and \citeyear{whitham74}, simplifying and improving upon results by \citealt{Chisnell55}, \citealt{Chisnell57}, and \citealt{Chester60}): 
\[\beta \rightarrow \left [2 + \left(2\gamma\over\gamma-1\right)^{1/2}\right]^{-1}. ~~~(n\rightarrow0)\]   
\citeauthor{whitham74} arrived at this form by reasoning that quantities just behind the shock front should evolve similarly to those found along a forward-traveling sound wave, for which there is an exact equation.   In general this is only a good approximation, because the shock moves more slowly than these forward characteristics and because conditions vary from one characteristic to another.  In the limit $n\rightarrow0$, however, there is no variation among characteristics, and \citeauthor{whitham74}'s approximation becomes exact.     

The isothermal limit $\gamma\rightarrow 1$ is characterized by \[\beta\rightarrow 0.~~~(\gamma\rightarrow1)\]  In this limit a strong shock is infinitely compressive and governed by the conservation of momentum.  The shock velocity is constant because the atmosphere above the shock front has negligible mass relative to the shell of post-shock material.  We note that  \citeauthor{whitham74}'s approximation is exact in this limit as well, because the shock no longer outruns forward characteristics. 

In the limit $n\rightarrow \infty$ the stellar structure is isothermal ($\gamma_p\rightarrow1$) and transitions from power-law to exponential in its depth dependence. It is reassuring, therefore, that equation (\ref{betaFit}) gives 
\[ \beta \rightarrow \left[1.78 +  \left(4.321 \gamma \over \gamma-1\right)^{0.469}\right]^{-1}  ~~~(n\rightarrow \infty)  \]
in this limit.  As \citet{1968JFM....32..305H} notes, this limit coincides with the case of an exponential atmosphere ($\gamma_p=1$); we reproduce his solutions and those of Raizer (1964).   Equation (\ref{betaFit}) demonstrates that the shock acceleration index is indistinguishable from the exponential case for all $n\gtrsim 10^2$ -- or in practical terms, any time that the distance to the surface is very far when measured relative to the density scale height. 

In the  $\gamma\rightarrow \infty$ limit of incompressible flow, $\beta$ takes the definite form $[A(n) + B(n)^{C(n)}]^{-1}$.   We know of no physical explanation for this result. 

We end this section by noting that for the specific case $\gamma=3/2$ and $n=5$, we find $\beta = 1/5$ and $\lambda=1$ (at least to within a part in $10^8$, while the fit of equation (\ref{betaFit}) gives $\lambda=1.01$).   This is unlikely to be a coincidence, although we have not identified any simplification in the dynamical equations for this case. 

\section{Post-Shock Flow and Asymptotic Free Expansion}
After each parcel of gas has been swept into motion by the shock, it continues to accelerate until its internal energy is spent and it has reached the terminal velocity $v_f(m)$.   To describe this we employ the Lagrangian method of \citet{1999ApJ...510..379M}.  This naturally provides quantities like $v_f(m)/v_s(m)$, and continues smoothly through the point of breakout; however it does not yield eigenvalues like $\beta$ as readily as \citeauthor{Sakurai60}'s method.   In order to correct a couple typos in  \citeauthor{1999ApJ...510..379M}'s Appendix (which do not affect their results), we write out the equation.  The Lagrangian self-similar time and space coordinates are $\eta = t/t_0(m)$ and $S=x/x_0(m)$, where $t_0(m)$ is the time at which the shock crosses $x_0(m)$.  For a given fluid element both $\eta$ and $S$ decline from unity to $-\infty$ as a fluid element accelerates outward; shock breakout (neglecting radiative effects) is at $\eta=0$, and the element exits the boundaries of the progenitor somewhat later, when $S=0$.  As  \citeauthor{1999ApJ...510..379M} discuss, the pressure and density distributions and the resulting acceleration can be computed from these variables, and the equating resulting fluid acceleration to $\ddot{x}(m)=S''(\eta) x_0(m)/t_0(m)^2$ yields
\begin{eqnarray}\label{eq:Lagrangian-accel} 
(1+&\lambda&)^2  S''(\eta)=-\frac{2}{\gamma+1}\left(\frac{\gamma-1}{\gamma+1}\right)^{\gamma} \times \nonumber \\
& & \left\lbrace  \frac{n-2\lambda}{\Sigma(\eta)^{\gamma}} - \gamma(\lambda+1)\eta \frac{\lambda S'(\eta) + (\lambda+1)\eta S''(\eta)}{\Sigma(\eta)^{\gamma+1}} \right\rbrace
\end{eqnarray}
where $\Sigma(\eta)\equiv S(\eta)-(\lambda+1)\eta S'(\eta)$.   Note that relative to \citeauthor{1999ApJ...510..379M}'s equation (A2) and its preceding discussion, all instances of $\lambda-1$ have been corrected to $\lambda+1$.  We integrate equation (\ref{eq:Lagrangian-accel}) from the post-shock conditions $S(1)=1$, $S'(1)=2/[(\lambda+1)(\gamma+1)]$ to a very large negative value of $\eta$ (typically $-\eta > 10^{20}$). 

\subsection{The acceleration factor and its limits} \label{S:VfVsLimits}

 We use the asymptotic form $[S_f - S'(\eta)] \propto (-\eta)^{-(\gamma-1)}$, valid for $\gamma>1$ (where $S'_f = S'(\eta\rightarrow\infty)$), to deduce $v_f(m)/v_s(m) = (\lambda+1) S'_f/S'(1) $.  We present this in Table \ref{Table:VfVs}; the approximation
\begin{eqnarray} \label{eq:vfvsApprox}
{v_f(m)\over v_s(m)} &=& {2\over \gamma+1}\left[ \left(2\gamma\over\gamma-1\right)^{D} + E\right]~~ \mbox{where}\\
D&=& \frac12 - {1.832 \over (1.083/n)^{0.828} + 1} + {1.655\over (1.5/n)^{0.883}+1}, \nonumber \\
E &=&1 -{0.803\over (2.39/n)^{1.18}+1} \nonumber 
\end{eqnarray} 
is reasonably accurate: over the values in Table \ref{Table:VfVs}, the rms error is 0.6\% and the maximum error is 1.8\%.    We do not have solutions for very small values of $\gamma-1$, however, so we cannot check accuracy in that regime. 

The form of equation (\ref{eq:vfvsApprox}) gives some insight into the nature of post-shock acceleration, because the initial factor $2/(\gamma+1)$ expresses the immediate post-shock velocity in units of $v_s$, so the factor $[2\gamma/(\gamma-1)]^D+E$ captures the subsequent acceleration from the post-shock state to the final state of free expansion. 

A clear limit of equation (\ref{eq:vfvsApprox}) is that in which the initial density distribution becomes uniform: 
\[ {v_f(m)\over v_s(m)} \rightarrow {2\over \gamma+1}\left[  \left(2\gamma\over\gamma-1\right)^{1/2} + 1\right]. ~~(n\rightarrow0)\] 
This limit is a consequence of the isentropic nature of the post-shock flow, which ensures that the Riemann invariant $v+2c_s/(\gamma-1)$ is conserved along outward-traveling sound fronts from the post-shock state to the freely expanding state; here $c_s$ is the adiabatic sound speed. 

In the opposite limit, corresponding to an exponential atmosphere, 
\[  {v_f(m)\over v_s(m)} \rightarrow {2\over \gamma+1}\left[  \left(2\gamma\over\gamma-1\right)^{1/3} + \frac15 \right]. ~~(n\rightarrow\infty)\] 
to within 3\%.
\citet{1968JFM....32..305H} and \citet{1966ApJ...143...48G} apply a Eulerian approach and do not provide this ratio.  
\citet{Raizer64} does computes post-shock acceleration in his Lagrangian treatment:  \citet{zeldovich67} report that he finds an increase of velocity, from the post-shock state to the time the shock blows out ($v_s\rightarrow\infty$) by the factor (1.54,1.85) for $\gamma=1.2,5/3$, respectively.  Furthermore,  \citeauthor{zeldovich67} report that \citeauthor{Raizer64} finds practically no subsequent acceleration.   According to equation (\ref{eq:vfvsApprox}), the ratio of final to post-shock acceleration is (2.43, 1.88) for the same values of $\gamma$: post-blowout acceleration is a strong function of $\gamma$. 

In the isothermal limit $\gamma\rightarrow 1$, approximation (\ref{eq:vfvsApprox}) and the data of Table 2 indicate that $v_f/v_s$ diverges proportionally to $(\gamma-1)^{-q}$ for $1/5\lesssim q\leq1/2$, as one might expect from the divergence of the internal energy.   
In the incompressible limit $\gamma\rightarrow \infty$, the final velocity limits to $2^{D(n)}+E(n)$ times the post-shock velocity, but the post-shock velocity becomes negligible compared to $v_s$; hence $v_f/v_s\rightarrow0$. 

One could describe the parameter dependence of other aspects of our planar shock emergence problem, such as the values of $S$ and $S'$ at $\eta=0$, which describe both the instant of breakout and the deep ejecta. Another example would be the coefficient in the relation $[v_f(m) - v(m,\eta)] \propto |S|^{-(\gamma-1)}$, which holds for $(-\eta)\gg1$ and from which one can express the influence of spherical expansion on the final velocity of fluid elements with finite values of $(|x_0|/R_*)^{\gamma-1}$ \citep{1992ApJ...400..192K,1999ApJ...510..379M}, where $R_*$ is the radius of curvature.  But since these quantities are readily available from the integration of equation (\ref{eq:Lagrangian-accel}) and are not as fundamental as  $\beta$ and $v_f/v_s$, we omit them.

\section{Discussion}\label{S:Discussion} 

We intend our study of the parameter dependence and limiting cases of the shock emergence problem to be useful for future investigations, and we envision several possibilities.   

First, real stellar structures are not perfectly characterized by ideal polytropes, and radial variations in the polytropic index near the stellar surface can influence the dynamics of shock emergence.   
\citet{1999ApJ...510..379M} propose  the shock velocity approximation
\begin{equation}\label{vs_MM99} 
v_s = \Lambda \left(E_{\rm in}\over m_{\rm ej} \right)^{1/2} \left(m_{\rm ej} \over \rho_0 r^3\right)^\beta 
\end{equation} 
for a spherical explosion of energy $E_{\rm in}$ traveling through a mass distribution described by radius $r$, density $\rho_0$, and enclosed ejecta mass $m_{\rm ej}$. {This form matches the properties of an interior blastwave (an energy-conserving flow in which $v_s(r)^2 m_{\rm ej}(r)\propto  E_{\rm in}$) with those of planar shock acceleration (a second-type similarity solution with eigenvalue indices)}. 
\citeauthor{1999ApJ...510..379M}  show that equation (\ref{vs_MM99}) achieves an accuracy of a few percent {in spherical, adiabatic supernova models} when $\Lambda$ and $\beta$ are taken from a representative Sedov blastwave and an $n=3$ polytropic atmosphere, respectively.   
To improve the accuracy of the energy scale for relativistic ejection,  \citet{2001ApJ...551..946T} adjust $\Lambda$ according to the moments of the density distribution interior {to $r$, while holding $\beta$ fixed.   }
The information provided in Table \ref{Table:beta} and equation (\ref{betaFit}) {should allow for further improvements in which $v_s$ better responds to the local conditions}.\footnote{It may also be possible to incorporate  \citet{1993PhFl....5.1035W}'s second-type similarity solutions for spherical blastwaves in steep density distributions ($d\ln\rho/d\ln r<-3$), although the rapid variation of $d\ln\rho/d\ln r$ in subsurface regions of stars poses a difficulty.} 

Second, for any hydrostatic envelope in which radiation pressure is initially negligible, there exists a range of shock strengths which are both strong and yet not dominated by radiation pressure in the post-shock state.   These might develop in the steepening of finite-amplitude sound pulses \citep{2004MNRAS.354.1053W} or in the explosion launched by the impact of a comet or asteroid on the atmosphere \citep{1994ApJ...429..863C}.   Insofar as the post-shock gas is described by a characteristic value of $\gamma$ different from 4/3, our results indicate how shock emergence is changed. 

Third, any investigation of the phenomena surrounding shock acceleration and shock emergence \citep[e.g., shock instability;][]{1994ApJ...435..815L} must use the ideal planar solution as its reference state. Understanding this solution's parameter dependence and limits therefore adds insight into the phenomenon under study. 

This work was supported by an NSERC Discovery Grant.  The authors are indebted to the referee 
and to Chris McKee 
for useful comments, and to Dylanne Dearborn and Roger Chevalier for help in our attempts to obtain \citet{Raizer64}.

\begin{deluxetable}{clllllllllllc} 
\tablecaption{Shock acceleration index $\beta$ \label{Table:beta}}
\tablehead{
\colhead{$n$} & 
\colhead{$\gamma=10^6$} &
\colhead{$\gamma=10^5$} &
\colhead{$\gamma=10^4$} &
\colhead{$\gamma=10^3$} &
\colhead{$\gamma=10^2$} &
\colhead{$\gamma=10$} & \colhead{$\gamma=5$} & \colhead{$\gamma=3$} &\colhead {$\gamma=2$} &\colhead{$\gamma=5/3$}  }
\startdata
$10^{-6}$ & 0.29289315 & 0.29289260 & 0.29288714 & 0.29283251 & 0.29228329 & 0.28647450 & 0.27924077 & 0.26794918 & 0.24999998 & 0.23606796 \\
$10^{-5}$ & 0.29289304 & 0.29289250 & 0.29288704 & 0.29283241 & 0.29228319 & 0.28647438 & 0.27924064 & 0.26794903 & 0.24999981 & 0.23606777 \\
$10^{-4}$ & 0.29289202 & 0.29289147 & 0.29288601 & 0.29283138 & 0.29228215 & 0.28647322 & 0.27923935 & 0.26794755 & 0.24999810 & 0.23606593 \\
$10^{-3}$ & 0.29288175 & 0.29288120 & 0.29287574 & 0.29282110 & 0.29227175 & 0.28646167 & 0.27922645 & 0.26793281 & 0.24998100 & 0.23604750 \\
$10^{-2}$ & 0.29277948 & 0.29277893 & 0.29277346 & 0.29271871 & 0.29216823 & 0.28634660 & 0.27909810 & 0.26778608 & 0.24981098 & 0.23586431 \\
$10^{-1}$ & 0.29179950 & 0.29179895 & 0.29179337 & 0.29173755 & 0.29117637 & 0.28524584 & 0.27787244 & 0.26638903 & 0.24819942 & 0.23413390 \\
1 & 0.28502240 & 0.28502178 & 0.28501554 & 0.28495308 & 0.28432550 & 0.27773019 & 0.26962150 & 0.25718336 & 0.23791739 & 0.22335005 \\
2 & 0.28096608 & 0.28096542 & 0.28095885 & 0.28089311 & 0.28023286 & 0.27331989 & 0.26488204 & 0.25205903 & 0.23244760 & 0.21779161 \\
3 & 0.27852654 & 0.27852586 & 0.27851912 & 0.27845169 & 0.27777456 & 0.27070121 & 0.26210529 & 0.24911283 & 0.22938205 & 0.21472787 \\
4 & 0.27690616 & 0.27690548 & 0.27689863 & 0.27683020 & 0.27614306 & 0.26897608 & 0.26029132 & 0.24721016 & 0.22743179 & 0.21279709 \\
5 & 0.27575433 & 0.27575364 & 0.27574673 & 0.27567764 & 0.27498397 & 0.26775669 & 0.25901642 & 0.24588314 & 0.22608481 & 0.21147160 \\
6 & 0.27489450 & 0.27489380 & 0.27488685 & 0.27481730 & 0.27411908 & 0.26685018 & 0.25807251 & 0.24490599 & 0.22509974 & 0.21050626 \\
7 & 0.27422857 & 0.27422787 & 0.27422088 & 0.27415099 & 0.27344944 & 0.26615027 & 0.25734597 & 0.24415692 & 0.22434841 & 0.20977222 \\
8 & 0.27369780 & 0.27369710 & 0.27369009 & 0.27361994 & 0.27291585 & 0.26559379 & 0.25676970 & 0.24356465 & 0.22375664 & 0.20919542 \\
9 & 0.27326496 & 0.27326425 & 0.27325722 & 0.27318687 & 0.27248079 & 0.26514085 & 0.25630156 & 0.24308471 & 0.22327860 & 0.20873031 \\
10 & 0.27290528 & 0.27290457 & 0.27289753 & 0.27282702 & 0.27211933 & 0.26476509 & 0.25591380 & 0.24268799 & 0.22288441 & 0.20834736 \\
$10^{2}$ & 0.26947121 & 0.26947050 & 0.26946332 & 0.26939150 & 0.26867090 & 0.26120469 & 0.25226650 & 0.23899134 & 0.21925258 & 0.20484213 \\
$10^{3}$ & 0.26906781 & 0.26906709 & 0.26905989 & 0.26898795 & 0.26826612 & 0.26078961 & 0.25184438 & 0.23856746 & 0.21884069 & 0.20444712 \\
$10^{4}$ & 0.26902677 & 0.26902605 & 0.26901886 & 0.26894690 & 0.26822495 & 0.26074742 & 0.25180152 & 0.23852446 & 0.21879895 & 0.20440713 \\
$10^{5}$ & 0.26902266 & 0.26902194 & 0.26901475 & 0.26894279 & 0.26822083 & 0.26074320 & 0.25179722 & 0.23852015 & 0.21879477 & 0.20440312 \\
$10^{6}$ & 0.26902225 & 0.26902153 & 0.26901434 & 0.26894238 & 0.26822041 & 0.26074278 & 0.25179679 & 0.23851972 & 0.21879436 & 0.20440272  \\
\enddata
\end{deluxetable}

\begin{deluxetable}{clllllllllll} 
\tablecaption{Shock acceleration factor $v_f/v_s$}

\tablehead{
\colhead{$n$} & 
\colhead{$\gamma=10$} & \colhead{$\gamma=5$} & \colhead{$\gamma=3$} &\colhead {$\gamma=2$} &\colhead{$\gamma=5/3$} & \colhead{$\gamma=3/2$} &
\colhead{$\gamma=7/5$} & 
 \colhead{$\gamma=4/3$} & \colhead{$\gamma=9/7$}    }  
\startdata
$10^{-6}$ & 0.45285582 & 0.86037753 & 1.36602118 & 1.99999136 & 2.42702649 & 2.75941695 & 3.03724468 & 3.27999086 & 3.49275349 \\
$10^{-5}$ & 0.45284794 & 0.86035969 & 1.36599636 & 1.99991857 & 2.42693295 & 2.75930702 & 3.03713155 & 3.27847912 & 3.49263769 \\
$10^{-4}$ & 0.45278311 & 0.86021382 & 1.36570632 & 1.99933390 & 2.42608170 & 2.75822288 & 3.03601982 & 3.27734354 & 3.49151902 \\
$10^{-3}$ & 0.45230353 & 0.85913741 & 1.36354654 & 1.99506907 & 2.41967881 & 2.74980213 & 3.02586515 & 3.26622378 & 3.47995904 \\
$10^{-2}$ & 0.44914536 & 0.85209731 & 1.34966085 & 1.96792794 & 2.37969643 & 2.69731006 & 2.96117950 & 3.19010855 & 3.39413426 \\
$10^{-1}$ & 0.43299067 & 0.81682384 & 1.28214995 & 1.84166283 & 2.19974497 & 2.46728053 & 2.68379133 & 2.86763036 & 3.02862713 \\
1 & 0.37506576 & 0.70203021 & 1.08820933 & 1.53116797 & 1.79844429 & 1.98896476 & 2.13717743 & 2.25915336 & 2.36311029 \\
2 & 0.34455376 & 0.64686880 & 1.00657057 & 1.42230689 & 1.67401293 & 1.85312095 & 1.99201909 & 2.10581684 & 2.20243857 \\
3 & 0.32711534 & 0.61597751 & 0.96228259 & 1.36626908 & 1.61234177 & 1.78782363 & 1.92386840 & 2.03523614 & 2.12966295 \\
4 & 0.31607498 & 0.59655629 & 0.93475556 & 1.33209762 & 1.57533069 & 1.74914052 & 1.88391867 & 1.99423027 & 2.08770216 \\
5 & 0.30851925 & 0.58330933 & 0.91608575 & 1.30913793 & 1.55065918 & 1.72353305 & 1.85762232 & 1.96737210 & 2.06033613 \\
6 & 0.30304299 & 0.57372626 & 0.90262297 & 1.29263311 & 1.53304634 & 1.70533006 & 1.83899322 & 1.94840504 & 2.04106258 \\
7 & 0.29889916 & 0.56648342 & 0.89246676 & 1.28029414 & 1.51985203 & 1.69172734 & 1.82510463 & 1.93429573 & 2.02675344 \\
8 & 0.29565625 & 0.56082202 & 0.88454049 & 1.27061632 & 1.50960199 & 1.68117857 & 1.81435616 & 1.92338863 & 2.01570684 \\
9 & 0.29305274 & 0.55627762 & 0.87818307 & 1.26289956 & 1.50140888 & 1.67263063 & 1.80579054 & 1.91470828 & 2.00691959 \\
10 & 0.29091665 & 0.55255066 & 0.87297252 & 1.25662285 & 1.49471099 & 1.66575381 & 1.79880288 & 1.90763458 & 1.99976839 \\
$10^{2}$ & 0.27174184 & 0.51913639 & 0.82607918 & 1.20029294 & 1.43475741 & 1.60437973 & 1.73661580 & 1.84484855 & 1.93645313 \\
$10^{3}$ & 0.26961772 & 0.51544773 & 0.81984467 & 1.19413475 & 1.42348349 & 1.59846719 & 1.72707902 & 1.84023222 & 1.93002834 \\
$10^{4}$ & 0.26940604 & 0.51486690 & 0.81768673 & 1.19303976 & 1.42366157 & 1.59710106 & 1.72484391 & 1.83940673 & 1.92736575 \\
$10^{5}$ & 0.26938403 & 0.51503932 & 0.81763521 & 1.19120321 & 1.42376162 & 1.59444347 & 1.72476060 & 1.83927471 & 1.92730526 \\
$10^{6}$ & 0.26898345 & 0.51486171 & 0.81763006 & 1.18946787 & 1.42168664 & 1.59284112 & 1.72163227 & 1.83940534 & 1.92594338
\enddata
\label{Table:VfVs}
\end{deluxetable}

\clearpage
\begin{turnpage}
\begin{deluxetable}{clllllllllcc} 
\tablenum{1}
\tablecaption{ --  Continued}
\tablehead{
\colhead{$n$} & 
\colhead{$\gamma=3/2$} &
\colhead{$\gamma=7/5$} &
\colhead{$\gamma=4/3$} &
\colhead{$\gamma=9/7$} &
\colhead{$\gamma=1.15$} &
\colhead{$\gamma=1.10$} & \colhead{$\gamma=1.06$} & \colhead{$\gamma=1.01$} &\colhead {$\gamma=1.001$} &\colhead{$\gamma=1+10^{-4}$} &\colhead{$\gamma=1+10^{-5}$}  }       
\startdata
$10^{-6}$ & 0.22474485 & 0.21525042 & 0.20710676 & 0.19999998 & 0.16903939 & 0.14946752 & 0.12587822 & 0.06168015 & 0.02139325 & 0.00697212 & 0.00222610 \\
$10^{-5}$ & 0.22474466 & 0.21525022 & 0.20710656 & 0.19999978 & 0.16903920 & 0.14946734 & 0.12587807 & 0.06168008 & 0.02139324 & 0.00697212 & 0.00222610 \\
$10^{-4}$ & 0.22474274 & 0.21524825 & 0.20710457 & 0.19999778 & 0.16903728 & 0.14946556 & 0.12587652 & 0.06167943 & 0.02139311 & 0.00697210 & 0.00222610 \\
$10^{-3}$ & 0.22472354 & 0.21522862 & 0.20708471 & 0.19997783 & 0.16901808 & 0.14944769 & 0.12586101 & 0.06167292 & 0.02139185 & 0.00697191 & 0.00222607 \\
$10^{-2}$ & 0.22453274 & 0.21503352 & 0.20688741 & 0.19977969 & 0.16882748 & 0.14927042 & 0.12570727 & 0.06160839 & 0.02137936 & 0.00697008 & 0.00222584 \\
$10^{-1}$ & 0.22273535 & 0.21319978 & 0.20503642 & 0.19792385 & 0.16705409 & 0.14762734 & 0.12428800 & 0.06101711 & 0.02126501 & 0.00695327 & 0.00222365 \\
1 & 0.21173503 & 0.20213822 & 0.19400295 & 0.18697155 & 0.15699241 & 0.13850153 & 0.11657586 & 0.05791760 & 0.02066532 & 0.00686464 & 0.00221214 \\
2 & 0.20619672 & 0.19667012 & 0.18862853 & 0.18170115 & 0.15236756 & 0.13440382 & 0.11319085 & 0.05659923 & 0.02040824 & 0.00682633 & 0.00220716 \\
3 & 0.20318002 & 0.19371804 & 0.18574712 & 0.17889128 & 0.14995044 & 0.13228214 & 0.11145283 & 0.05592780 & 0.02027633 & 0.00680657 & 0.00220459 \\
4 & 0.20129129 & 0.19187868 & 0.18395846 & 0.17715216 & 0.14846962 & 0.13098826 & 0.11039710 & 0.05552103 & 0.02019600 & 0.00679450 & 0.00220302 \\
5 & 0.20000000 & 0.19062491 & 0.18274204 & 0.17597158 & 0.14747055 & 0.13011769 & 0.10968834 & 0.05524823 & 0.02014191 & 0.00678635 & 0.00220197 \\
6 & 0.19906222 & 0.18971624 & 0.18186180 & 0.17511833 & 0.14675145 & 0.12949217 & 0.10917982 & 0.05505258 & 0.02010302 & 0.00678049 & 0.00220120 \\
7 & 0.19835060 & 0.18902773 & 0.18119558 & 0.17447309 & 0.14620925 & 0.12902113 & 0.10879727 & 0.05490541 & 0.02007369 & 0.00677606 & 0.00220063 \\
8 & 0.19779228 & 0.18848814 & 0.18067391 & 0.17396818 & 0.14578590 & 0.12865367 & 0.10849906 & 0.05479069 & 0.02005080 & 0.00677260 & 0.00220018 \\
9 & 0.19734262 & 0.18805395 & 0.18025441 & 0.17356237 & 0.14544622 & 0.12835905 & 0.10826008 & 0.05469876 & 0.02003242 & 0.00676981 & 0.00219982 \\
10 & 0.19697276 & 0.18769706 & 0.17990977 & 0.17322912 & 0.14516765 & 0.12811757 & 0.10806430 & 0.05462344 & 0.02001735 & 0.00676753 & 0.00219952 \\
$10^{2}$ & 0.19360194 & 0.18445447 & 0.17678575 & 0.17021368 & 0.14266166 & 0.12595038 & 0.10631033 & 0.05394810 & 0.01988142 & 0.00674689 & 0.00219684 \\
$10^{3}$ & 0.19322367 & 0.18409167 & 0.17643698 & 0.16987761 & 0.14238391 & 0.12571071 & 0.10611667 & 0.05387344 & 0.01986630 & 0.00674458 & 0.00219654 \\
$10^{4}$ & 0.19318538 & 0.18405496 & 0.17640170 & 0.16984362 & 0.14235584 & 0.12568649 & 0.10609710 & 0.05386590 & 0.01986477 & 0.00674435 & 0.00219651 \\
$10^{5}$ & 0.19318155 & 0.18405129 & 0.17639817 & 0.16984022 & 0.14235303 & 0.12568406 & 0.10609514 & 0.05386514 & 0.01986461 & 0.00674433 & 0.00219650 \\
$10^{6}$ & 0.19318116 & 0.18405092 & 0.17639782 & 0.16983988 & 0.14235275 & 0.12568382 & 0.10609495 & 0.05386507 & 0.01986460 & 0.00674432 & 0.00219650 \\
\enddata
\end{deluxetable}
\clearpage
\end{turnpage}

\end{document}